\documentclass[prl,twocolumn,twoside,preprintnumbers,superscriptaddress,nofootinbib]{revtex4}
\usepackage{amsmath,slashed}
\usepackage{graphicx,graphics}
\usepackage{dcolumn}
\usepackage[hyperfootnotes=false]{hyperref}
\usepackage{xspace}

\newcommand{\Fpi}{F_\pi}
\newcommand{\mpi}{M_{\pi}}
\newcommand{\mpii}{M_{\pi^0}}
\newcommand{\Order}{\mathcal{O}}
\newcommand{\MeV}{\,\text{MeV}}
\newcommand{\GeV}{\,\text{GeV}}

\newcommand{\beq}{\begin{equation}}
\newcommand{\eeq}{\end{equation}}
\newcommand{\diff}{\text{d}}
\newcommand{\sis}{s_\text{is}}
\newcommand{\siv}{s_\text{iv}}
\newcommand{\sthr}{s_\text{thr}}
\newcommand{\sm}{s_\text{m}}
\newcommand{\grhop}{g_\text{eff}}
\newcommand{\Mrhop}{M_\text{eff}}
\newcommand{\eps}{\epsilon}

\renewcommand{\Im}{\text{Im}\,}

\begin{document}

\preprint{INT-PUB-18-016}
\title{Pion-pole contribution to hadronic light-by-light scattering \\[1mm] in the anomalous magnetic moment of the muon}

\author{Martin Hoferichter}
\affiliation{Institute for Nuclear Theory, University of Washington, Seattle, WA 98195-1550, USA}
\author{Bai-Long Hoid}
\affiliation{Helmholtz-Institut f\"ur Strahlen- und Kernphysik (Theory) and \\Bethe Center for Theoretical Physics, University of Bonn, 53115 Bonn, Germany}
\author{Bastian Kubis}
\affiliation{Helmholtz-Institut f\"ur Strahlen- und Kernphysik (Theory) and \\Bethe Center for Theoretical Physics, University of Bonn, 53115 Bonn, Germany}
\author{Stefan Leupold}
\affiliation{Institutionen f\"or fysik och astronomi, Uppsala Universitet, Box 516, 75120 Uppsala, Sweden}
\author{Sebastian P.\ Schneider}
\affiliation{Helmholtz-Institut f\"ur Strahlen- und Kernphysik (Theory) and \\Bethe Center for Theoretical Physics, University of Bonn, 53115 Bonn, Germany}

\begin{abstract}
The $\pi^0$ pole constitutes the lowest-lying singularity of the hadronic light-by-light (HLbL) tensor, and thus, it provides the leading contribution in a dispersive approach 
to HLbL scattering in the anomalous magnetic moment of the muon $(g-2)_\mu$. It is unambiguously defined in terms of the doubly-virtual pion transition form factor, which in principle can be 
accessed in its entirety by experiment. We demonstrate that, in the absence of a direct measurement, the full space-like doubly-virtual form factor can be reconstructed very accurately
based on existing data for $e^+e^-\to 3\pi$, $e^+e^-\to e^+e^-\pi^0$, and the $\pi^0\to\gamma\gamma$ decay width. We derive a representation that incorporates 
all the low-lying singularities of the form factor, matches correctly onto the asymptotic behavior expected from perturbative QCD, and is suitable for the evaluation
of the $(g-2)_\mu$ loop integral. The resulting value, $a_\mu^{\pi^0\text{-pole}}=62.6^{+3.0}_{-2.5}\times 10^{-11}$, for the first time, represents a complete data-driven determination
of the pion-pole contribution with fully controlled uncertainty estimates. In particular, we show that already improved singly-virtual measurements alone would allow one to further reduce 
the uncertainty in $a_\mu^{\pi^0\text{-pole}}$.
\end{abstract}

\maketitle

\section{Introduction}

The anomalous magnetic moment of the muon $(g-2)_\mu$ constitutes a highly sensitive 
probe of physics beyond the Standard Model (BSM). Experimentally, its value is dominated
by the final report of the BNL E821 experiment~\cite{Bennett:2006fi}, which departs
from the SM by about $3\sigma$ and thus adds to the intriguing hints for
BSM physics observed in the muon sector. Currently, experiment and theory 
are known at the same level, so upcoming improved measurements at Fermilab~\cite{Grange:2015fou}
and, potentially, at J-PARC~\cite{Saito:2012zz} (see also~\cite{Gorringe:2015cma}) 
demand concurrent advances in the SM prediction. If the present discrepancy were to be confirmed, 
its significance would crucially depend on the theory uncertainties.   

\begin{figure}[t]
 \centering
\includegraphics[width=\linewidth]{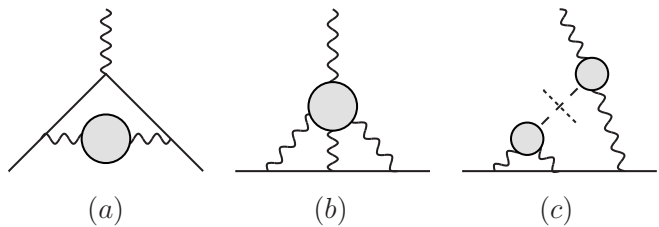}
 \caption{Diagrammatic representation of $(a)$ HVP, $(b)$ HLbL, and $(c)$ the $\pi^0$ pole in HLbL. Solid, dashed, and wiggly lines denote muons, pions, and photons, respectively, and the short-dashed line indicates that the pion is to be taken on-shell.}
 \label{fig:diagrams}
\end{figure}

In practice, this program requires improvements in two classes of hadronic corrections, see diagrams $(a)$ and $(b)$ in Fig.~\ref{fig:diagrams}:
hadronic vacuum polarization (HVP) and HLbL scattering~\cite{Jegerlehner:2009ry},
with QED at the five-loop order~\cite{Aoyama:2012wk,Aoyama:2014sxa,Aoyama:2017uqe} (and analytical cross checks at four loops~\cite{Kurz:2015bia,Kurz:2016bau,Laporta:2017okg}), 
electroweak corrections~\cite{Czarnecki:2002nt,Gnendiger:2013pva}, and higher-order iterations of HVP~\cite{Calmet:1976kd,Kurz:2014wya} and HLbL scattering~\cite{Colangelo:2014qya}
already known sufficiently accurately. 
As early as~\cite{Bouchiat:1961} it was realized that analyticity and unitarity 
allow one to express HVP in terms of the cross section $\sigma(e^+e^-\to\text{hadrons})$, so that improved measurements of the hadronic input directly benefit the SM prediction for $(g-2)_\mu$. 
Indeed, the most recent compilations~\cite{Jegerlehner:2017lbd,Davier:2017zfy,Keshavarzi:2018mgv}
already quote an uncertainty at the same level as or below HLbL scattering, even though tensions particularly in the $2\pi$ channel need to be resolved.
A similar approach for HLbL scattering, aiming at reconstructing the four-point function from its singularities, has only recently been developed~\cite{Hoferichter:2013ama,Colangelo:2014dfa,Colangelo:2014pva,Colangelo:2015ama,Colangelo:2017qdm,Colangelo:2017fiz}. The simplest such singularities take the form
of single-particle poles from the exchange of pseudoscalar mesons, with residues uniquely determined by transition form factors that are accessible in experiment, see diagram $(c)$ in Fig.~\ref{fig:diagrams}.  
In the context of hadronic models there has been wide agreement that the pion pole emerges as the numerically dominant contribution~\cite{deRafael:1993za,Bijnens:1995cc,Bijnens:1995xf,Bijnens:2001cq,Hayakawa:1995ps,Hayakawa:1996ki,Hayakawa:1997rq,Knecht:2001qg,Knecht:2001qf,Blokland:2001pb,RamseyMusolf:2002cy,Melnikov:2003xd,Masjuan:2012wy,Roig:2014uja,Benayoun:2014tra}, but its precise definition has been under debate, with variants such as a constant form factor at one vertex~\cite{Melnikov:2003xd} or even an off-shell pion~\cite{Jegerlehner:2009ry} being introduced.  

In the dispersive reconstruction of the HLbL tensor such ambiguities do not arise.  It is the form as derived in~\cite{Knecht:2001qg,Knecht:2001qf} that reproduces the correct residue (for a vector-meson-dominance transition form factor, this conclusion has also been confirmed using a dispersion relation for the Pauli form factor~\cite{Pauk:2014rfa}). As a consequence, most recent model-based evaluations have adopted this dispersive definition of pseudoscalar poles~\cite{Nyffeler:2016gnb,Masjuan:2017tvw,Guevara:2018rhj}, as have calculations of the pion transition form factor within lattice QCD~\cite{Gerardin:2016cqj} (complementary to lattice-QCD calculations aiming at the full HLbL contribution~\cite{Blum:2014oka,Green:2015sra,Blum:2015gfa,Blum:2016lnc,Blum:2017cer}).   
From an experimental point of view the pion pole can be considered similar to HVP, since a measurement of $e^+e^-\to e^+e^-\pi^0$ cross sections for space-like doubly-virtual kinematics fully determines 
its contribution to $(g-2)_\mu$. In this Letter we show that  
existing data for $e^+e^-\to 3\pi$, the space-like singly-virtual process $e^+e^-\to e^+e^-\pi^0$, and the $\pi^0\to\gamma\gamma$ decay width already severely constrain the doubly-virtual form factor again by virtue of analyticity and unitarity.
This calculation completes a dedicated effort to obtain a fully data-driven determination of the pion-pole
contribution to HLbL scattering~\cite{Niecknig:2012sj,Schneider:2012ez,Hoferichter:2012pm,Hoferichter:2014vra,Hoferichter:2017ftn,Hoferichter:2018kwz}, including a comprehensive error analysis and matching to short-distance constraints, with conceptual advances that will be invaluable for a similar approach to the $\eta$ and $\eta'$ poles~\cite{Stollenwerk:2011zz,Hanhart:2013vba,Kubis:2015sga,Xiao:2015uva}.

\section{Pion-pole contribution to $\boldsymbol{(g-2)_\mu}$}

A dispersive approach to HLbL scattering is based on a decomposition of the tensor describing the HLbL process $\gamma^*(q_1)\gamma^*(q_2)\to\gamma^*(-q_3)\gamma(q_4)$ into suitable scalar functions $\Pi_i$. In terms of these functions the contribution to $(g-2)_\mu$ may be written as~\cite{Colangelo:2017fiz}
\begin{align}
\label{amuHLbL}
a_\mu^\text{HLbL} &= \frac{\alpha^3}{432\pi^2} \int_0^\infty \diff\Sigma\, \Sigma^3 \int_0^1 \diff r\, r\sqrt{1-r^2} \int_0^{2\pi} \diff\phi \notag\\
&\qquad\times\sum_{i=1}^{12} T_i(\{q_j^2\}) \bar\Pi_i(\{q_j^2\}),
\end{align}
where the $\bar \Pi_i$ denote linear combinations of the full set of $\Pi_i$, $\alpha=e^2/(4\pi)$, the $T_i$ are known kernel functions, and the Euclidean virtualities $Q_i^2=-q_i^2$ are parameterized in terms of a single momentum scale $\Sigma$ according to~\cite{Eichmann:2015nra} 
\begin{align}
Q_{1,2}^2 &= \frac{\Sigma}{3} \Big( 1 - \frac{r}{2} \cos\phi \mp \frac{r}{2}\sqrt{3} \sin\phi \Big), \notag\\
Q_3^2 &= \frac{\Sigma}{3} \big( 1 + r \cos\phi \big).
\end{align}
The $\bar \Pi_i$ are then to be reconstructed in terms of their singularities, most prominently the pole originating from $\pi^0$ intermediate states. This singularity only appears in two of the $\bar \Pi_i$, 
\beq
\label{Pi1_pipole}
\bar \Pi_1^{\pi^0\text{-pole}}(\{q_i^2\})=\frac{F_{\pi^0\gamma^*\gamma^*}(q_1^2,q_2^2)F_{\pi^0\gamma^*\gamma^*}(q_3^2,0)}{q_3^2-\mpii^2},
\eeq
and $q_2^2\leftrightarrow q_3^2$ for $\bar \Pi_2$, with residues determined by the pion transition form factor $F_{\pi^0\gamma^*\gamma^*}(q_1^2,q_2^2)$ (the resulting representation is then equivalent to~\cite{Knecht:2001qf}). This form factor, defined by the matrix element
of two electromagnetic currents $j_\mu(x)$
\begin{align}
  \label{eq:defpiTFF}
  & i\int \diff^4x \, e^{iq_1\cdot x} \, \langle 0 \vert T \, j_\mu(x) \, j_\nu(0) \vert \pi^0(q_1+q_2) \rangle
  \notag \\
  & = -\epsilon_{\mu\nu\alpha\beta} \, q_1^\alpha \, q_2^\beta \, F_{\pi^0\gamma^*\gamma^*}(q_1^2,q_2^2),
\end{align}
is thus the central object of this Letter. Its normalization $F_{\pi\gamma\gamma}$ is related to the pion decay constant $\Fpi=92.28(9)\MeV$~\cite{Olive:2016xmw}  by a low-energy theorem~\cite{Adler:1969gk,Bell:1969ts,Bardeen:1969md}
\beq
\label{LET}
F_{\pi\gamma\gamma}=\frac{1}{4\pi^2\Fpi}.
\eeq
Experimentally, this low-energy theorem has been confirmed at the level of $1.4\%$ in the $\pi^0\to\gamma\gamma$ decay~\cite{Larin:2010kq}, which is the uncertainty that we will attach to the central value~\eqref{LET} in this Letter. We note that this accuracy is likely to improve by a factor of $2$ at PrimEx-II~\cite{Gasparian:2016oyl}, although
the role of chiral corrections remains to be understood~\cite{Kampf:2009tk}. 
While there is experimental information for space-like singly-virtual kinematics~\cite{Behrend:1990sr,Gronberg:1997fj,Aubert:2009mc,Uehara:2012ag}, 
the available data sets cover primarily large virtualities beyond the low-energy region $\lesssim 1\GeV$, the latter being most relevant for $(g-2)_\mu$, 
and doubly-virtual input is lacking completely. 

In the following, we construct a form factor representation
\beq
\label{TFF_final}
F_{\pi^0\gamma^*\gamma^*}=F_{\pi^0\gamma^*\gamma^*}^\text{disp}+F_{\pi^0\gamma^*\gamma^*}^\text{eff}+F_{\pi^0\gamma^*\gamma^*}^\text{asym}
\eeq
that proves to be remarkably universal: via the first, dispersive, term, it reproduces all low-energy singularities;
the second, small, term parameterizes higher intermediate states and high-energy contributions of the leading channels 
that are required to both fully saturate a sum rule for $F_{\pi\gamma\gamma}$ and to describe high-energy singly-virtual data;
and the third term, based on a pion distribution amplitude, makes the representation
respect all asymptotic constraints at $\Order(1/Q^2)$.

We begin by demonstrating how the crucial low-energy behavior of $F_{\pi^0\gamma^*\gamma^*}(q_1^2,q_2^2)$ can be reconstructed from its $2\pi$ and $3\pi$ singularities, which in turn can be extracted from $e^+e^-\to 2\pi,3\pi$.
The effective contribution is then added to enforce the correct form
factor normalization and implement the constraints from large-$Q^2$
data. Finally, the total contribution is smoothly matched to the last
term that ensures the asymptotic behavior expected from perturbative QCD.

\section{Low-energy properties}

For convenience, we decompose the form factor into components of definite isospin
\beq
\label{Fvs_isospin}
F_{\pi^0\gamma^*\gamma^*}(q_1^2,q_2^2)=F_{vs}(q_1^2,q_2^2)+F_{vs}(q_2^2,q_1^2),
\eeq
where the first (second) argument in $F_{vs}(q_1^2,q_2^2)$ denotes isovector (isoscalar) virtualities. 
Assuming the principle of maximum analyticity, this object fulfills a dispersion relation~\cite{Hoferichter:2014vra} (anomalous thresholds do not arise~\cite{Lucha:2006vc,Colangelo:2015ama})
\beq
\label{Fvs}
F_{vs}(q_1^2,q_2^2)=
\frac{1}{12\pi^2}\int^{\infty}_{4\mpi^2}\diff x\frac{q_{\pi}^3(x)\big(F_\pi^{V}(x)\big)^{*}f_1(x,q_2^2)}{x^{1/2}(x-q_1^2-i\eps)},
\eeq
where $q_\pi(s)=\sqrt{s/4-\mpi^2}$, $F_\pi^{V}(s)$ is the electromagnetic form factor of the pion, and $f_1(s,q^2)$ the partial-wave amplitude for $\gamma^*(q)\pi\to\pi\pi$. 
In practice, such an unsubtracted dispersion relation does not fully reproduce the low-energy theorem, the sum rule for $F_{vs}(0,0)=F_{\pi\gamma\gamma}/2$ being violated by about $10\%$~\cite{Hoferichter:2014vra}, which can be remedied by invoking a subtracted variant instead. 
However, for a subtracted dispersion relation the asymptotic behavior is only correct within the uncertainties within which the sum rule is fulfilled, so that,
when scanning over the range of permissible input variants, a form factor would be generated that behaves as a constant asymptotically, 
rendering the $(g-2)_\mu$ integral divergent. Therefore, we retain 
the unsubtracted version, introduce a cutoff $\siv$ in the integral, and restore the sum rule by adding an effective pole as explained below.

\begin{figure}[t]
 \centering
\includegraphics*[width=\linewidth]{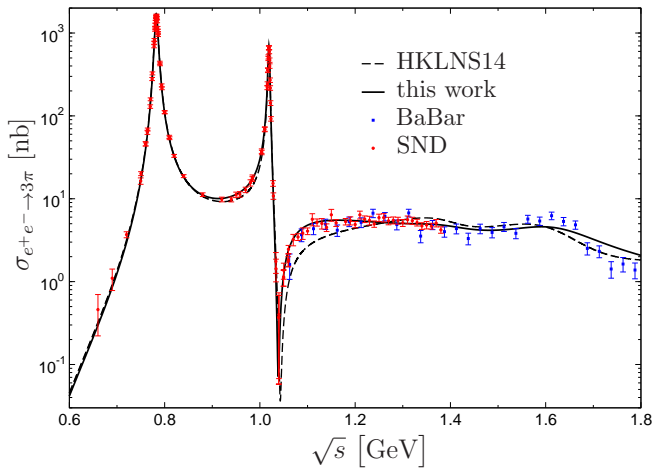}
 \caption{Fit to the $e^+e^-\to 3\pi$ cross section from SND~\cite{Achasov:2002ud,Achasov:2003ir} and BaBar~\cite{Aubert:2004kj}, in comparison to~\cite{Hoferichter:2014vra} (HKLNS14).}
 \label{fig:fit3pi}
\end{figure}

The integrand in~\eqref{Fvs} is reconstructed as follows: $F_\pi^{V}(s)$ is described by an Omn\`es representation~\cite{Omnes:1958hv} fit to the data from~\cite{Fujikawa:2008ma}. For the $\pi\pi$ $P$-wave phase shift we use~\cite{Colangelo:2001df,Caprini:2011ky},
a variant thereof that includes $\rho'$, $\rho''$ in an elastic approximation~\cite{Schneider:2012ez},
and the parameterization from~\cite{GarciaMartin:2011cn}.    
The main complexity, however, is hidden in $f_1(s,q^2)$, whose determination requires the solution of Khuri--Treiman equations~\cite{Khuri:1960zz} for $\gamma^*\pi\to\pi\pi$~\cite{Hoferichter:2014vra}.
These equations then predict the energy dependence of the partial wave, but the normalization $a(q^2)$ needs to be determined from experiment. At $q^2=M_\omega^2,M_\phi^2$ this normalization can be extracted from the $\omega,\phi\to3\pi$ decay widths, at $q^2=0$ its value is determined by another low-energy theorem~\cite{Wess:1971yu,Adler:1971nq,Terentev:1971cso,Aviv:1971hq,Witten:1983tw}, and for $q^2>9\mpi^2$ it is accessible in $e^+e^-\to 3\pi$. Building upon the parameterization derived in~\cite{Hoferichter:2014vra}, we include a conformal polynomial to better account for inelastic states above $1\GeV$ (including left-hand cuts beyond pion poles in the solution of the Khuri--Treiman equations), which indeed improves the fit significantly in particular above the $\phi$ and allows us to describe the SND~\cite{Achasov:2002ud,Achasov:2003ir} and low-energy BaBar~\cite{Aubert:2004kj} data with a reduced $\chi^2/\text{dof}\sim 1$, see Fig.~\ref{fig:fit3pi}. For more details of the representation we refer to~\cite{Hoferichter:2018kwz}.

In this way, \eqref{Fvs} determines, in principle, the full doubly-virtual form factor, but for the analytic continuation into the space-like region it is advantageous to apply yet another dispersion relation in the isoscalar variable
\beq
F_{vs}(-Q_1^2,q_2^2)=\frac{1}{\pi}\int_{\sthr}^{\sis} \diff y\frac{\Im F_{vs}(-Q_1^2,y)}{y-q_2^2-i\eps},
\eeq
with an isoscalar cutoff $\sis$ and an integration threshold $\sthr$. 
In the absence of electromagnetic effects and in the isospin limit, this representation
emerges with $\sthr=9\mpi^2$ because, for fixed $q_1^2<0$, the integral~\eqref{Fvs} has a nonzero
imaginary part if the phases of $F_\pi^V$ and of the partial wave $f_1$ differ, i.e.\ for $q_2^2> 9 \mpi^2$. 
In addition, we take into account the $\omega\to\pi^0\gamma$ decay in $a(q^2)$~\cite{Hoferichter:2014vra},
which results in $\sthr=\mpii^2$. This amounts to a double-spectral representation
\begin{align}
\label{low_energy}
F_{vs}^\text{disp}(-Q_1^2,-Q_2^2)&=
\frac{1}{\pi^2} \int_{4M_\pi^2}^{\siv} \diff x \int_{\sthr}^{\sis}  \frac{\diff y  \, \rho(x,y)}{\big(x+Q_1^2\big)\big(y+Q_2^2\big)},\notag\\
\rho(x,y)&=\frac{q_\pi^3(x)}{12\pi\sqrt{x}}\Im \Big[\big(F_\pi^{V}(x)\big)^*f_1(x,y)\Big]
\end{align}
to describe the low-energy properties of the form factor.

\section{Asymptotic behavior}

The representation~\eqref{low_energy} incorporates all of the low-lying singularities in the $2\pi$ and $3\pi$ channels.
When concentrating on the low-energy region, the effect of higher intermediate states can be efficiently suppressed by introducing subtractions~\cite{Hoferichter:2014vra}, 
but the resulting high-energy behavior disfavors such a representation for the $(g-2)_\mu$ application.   
Asymptotically, the transition form factor should fulfill~\cite{Lepage:1979zb,Lepage:1980fj,Brodsky:1981rp}
\beq
\label{pQCD}
 F_{\pi^0\gamma^*\gamma^*}(q_1^2,q_2^2)=-\frac{2\Fpi}{3}\int_0^1\diff x\frac{\phi_\pi(x)}{x q_1^2+(1-x) q_2^2}
+\Order\bigg(\!\frac{1}{q_i^{4}}\!\bigg),
\eeq
with a pion distribution amplitude $\phi_\pi(x)=6x(1-x)$, but a rigorous derivation based on the operator product expansion only exists for $|(q_1^2-q_2^2)/(q_1^2+q_2^2)|<1/2$~\cite{Manohar:1990hu}. 
Indeed, variants of~\eqref{pQCD} have been proposed including $\alpha_s$ corrections~\cite{delAguila:1981nk,Braaten:1982yp}, 
subleading terms in $1/Q^2$ and the Gegenbauer expansion of $\phi_\pi(x)$~\cite{Chernyak:1981zz,Chernyak:1983ej,Novikov:1983jt} within light-cone sum rules~\cite{Khodjamirian:1997tk,Agaev:2010aq,Mikhailov:2016klg}, Dyson--Schwinger equations~\cite{Raya:2015gva,Eichmann:2017wil}, and Regge theory~\cite{RuizArriola:2006jge,Arriola:2010aq,Gorchtein:2011vf}.
We implement these constraints as follows. First, as observed in~\cite{Khodjamirian:1997tk}, for space-like virtualities~\eqref{pQCD} can be expressed as a dispersion relation by a simple change of variables, formally represented by a (singular) double-spectral density
\beq
\rho^\text{asym}(x,y)=-2\pi^2\Fpi x y\delta''(x-y).
\eeq
In particular, introducing a lower matching point $\sm$, this defines an asymptotic contribution
\beq
\label{asym}
F_{\pi^0\gamma^*\gamma^*}^\text{asym}(q_1^2,q_2^2)
 = 2\Fpi\int_{\sm}^\infty \diff x \frac{q_1^2q_2^2}{(x-q_1^2)^2(x-q_2^2)^2}
\eeq
that ensures consistency with~\eqref{pQCD} for nonvanishing virtualities.

\begin{figure}[t]
 \centering
\includegraphics[width=\linewidth,clip]{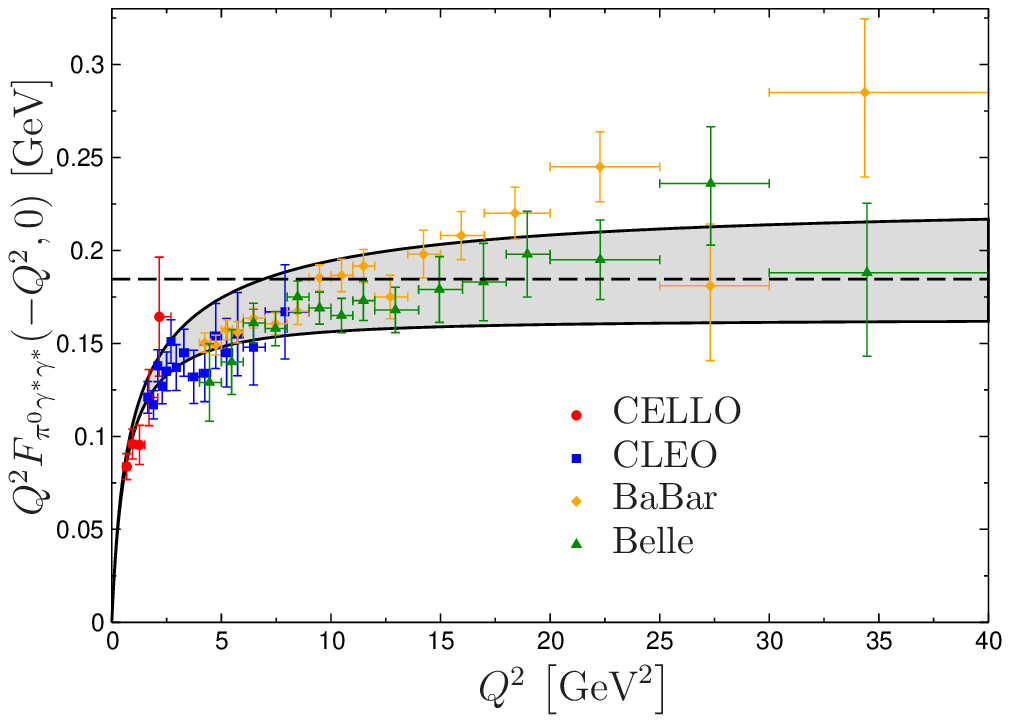}\\
\includegraphics[width=\linewidth,clip]{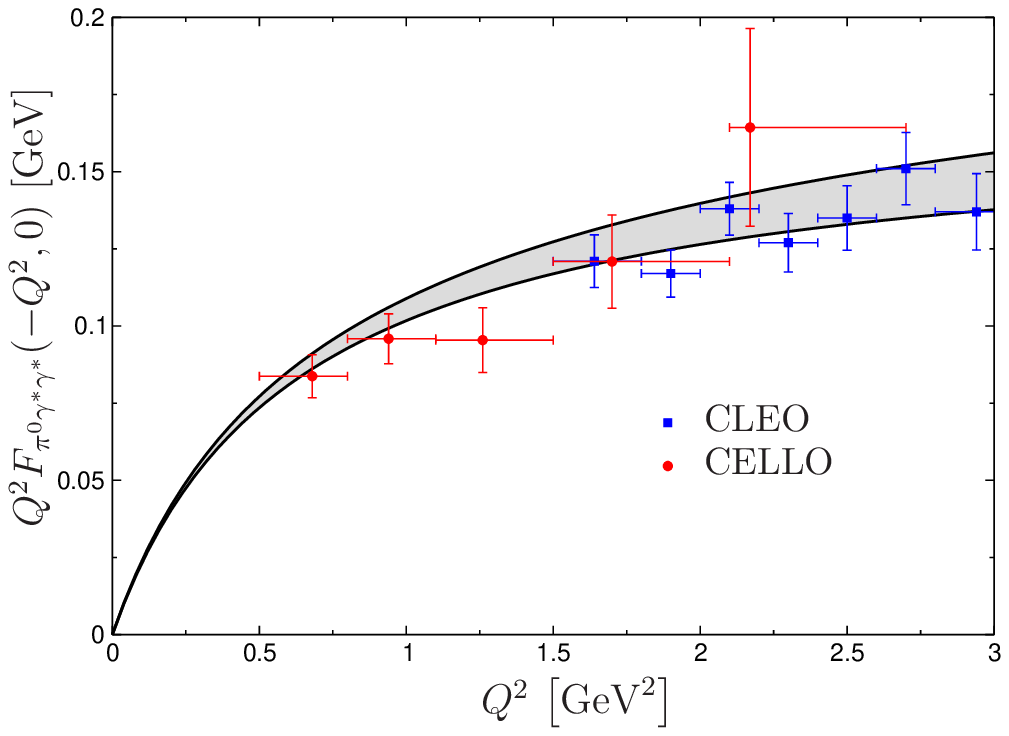}
 \caption{Resulting uncertainty band for the asymptotic space-like singly-virtual transition form factor (upper) and the corresponding prediction for the low-energy region (lower). The dashed line refers to the BL limit $2\Fpi$.}
 \label{fig:TFF_spacelike}
\end{figure}

For the singly-virtual asymptotics the situation is more complicated, given that now the low-energy part~\eqref{low_energy} contributes and~\eqref{pQCD} becomes less reliable. 
However, our representation~\eqref{low_energy} already saturates the Brodsky--Lepage (BL) limit $\lim_{Q^2\to\infty}Q^2F_{\pi^0\gamma^*\gamma^*}(-Q^2,0)=2\Fpi$ at the level of $55\%$, so that only the remainder needs to be 
generated by higher intermediate states and high-energy contributions, which can be conveniently achieved by an effective pole in the double-spectral density. Such a pole amounts to an additional term
\beq
F_{\pi^0\gamma^*\gamma^*}^\text{eff}(q_1^2,q_2^2)=\frac{\grhop}{4\pi^2\Fpi}\frac{\Mrhop^4}{(\Mrhop^2-q_1^2)(\Mrhop^2-q_2^2)},
\eeq
where the residue $\grhop$ is adjusted to restore the sum rule for $F_{\pi\gamma\gamma}$ and the mass parameter $\Mrhop$ defines the asymptotic value in the singly-virtual direction without affecting the doubly-virtual behavior at $\Order(1/Q^2)$. 
The resulting parameters $\grhop$ and $\Mrhop$ end up around $10\%$ and $1.5$--$2\GeV$, respectively, consistent with the interpretation as an effective pole that subsumes the effects of higher intermediate states and high-energy contributions.

For the numerical analysis, we vary $\grhop$ within the range defined by the PrimEx experiment~\cite{Larin:2010kq} and fit $\Mrhop$ to the space-like data~\cite{Behrend:1990sr,Gronberg:1997fj,Aubert:2009mc,Uehara:2012ag}. In our formalism, the tension of the BaBar data~\cite{Aubert:2009mc} both with the BL limit
and the other data sets manifests itself in a strong sensitivity of the fit results on the lower threshold above which data are included as well as in
a deteriorating $\chi^2$ once that threshold is increased. For that reason, we define our central value by the fit to all data sets but BaBar, leading to an asymptotic value almost exactly at the BL limit.
However, we also consider fit variants ranging from fits including the BaBar data set to imposing the strict BL limit. The envelope of the $3\sigma$ bands for each of these
variants corresponds to a variation by $^{+20}_{-10}\%$ around the asymptotic value in our preferred fit, which then defines the error band in Fig.~\ref{fig:TFF_spacelike}.  
Since only data above $5\GeV^2$ have been included in the fit,
this also defines the prediction for the low-energy region, improving the corresponding result from~\cite{Hoferichter:2014vra} by the matching to the asymptotic constraints. 
In particular, we find for the slope parameter
\begin{align}
a_\pi&=\frac{\mpii^2}{F_{\pi\gamma\gamma}}\frac{\partial}{\partial q^2}F_{\pi^0\gamma^*\gamma^*}(q^2,0)\bigg|_{q^2=0}\notag\\
&=31.5(2)_{F_{\pi\gamma\gamma}}(8)_\text{disp}(3)_\text{BL}\times 10^{-3}\notag\\
&=31.5(9)\times 10^{-3}.
\end{align}
The upward shift compared to $a_\pi=30.7(6)\times 10^{-3}$~\cite{Hoferichter:2014vra} precisely reflects the matching to the asymptotic region, see~\cite{Hoferichter:2018kwz} for details.

\section{Results for $\boldsymbol{(g-2)_\mu}$}

The final representation~\eqref{TFF_final} then defines the pion-pole contribution to HLbL scattering by means of~\eqref{amuHLbL} and~\eqref{Pi1_pipole}
\begin{align}
\label{result_final}
 a_\mu^{\pi^0\text{-pole}}&=62.6(1.7)_{F_{\pi\gamma\gamma}}(1.1)_\text{disp}(^{2.2}_{1.4})_\text{BL}(0.5)_\text{asym}\times 10^{-11}\notag\\
 &=62.6^{+3.0}_{-2.5}\times 10^{-11}.
\end{align}
One major component in the error analysis is the uncertainty in $F_{\pi\gamma\gamma}$, currently known at $1.4\%$ from PrimEx~\cite{Larin:2010kq},
but already the preliminary PrimEx-II update of $0.85\%$~\cite{Gasparian:2016oyl} reduces the $F_{\pi\gamma\gamma}$-related and total error to $1.0$ and $^{+2.7}_{-2.1}\times 10^{-11}$, respectively.
The uncertainty in the dispersive part~\eqref{low_energy} is estimated by varying the cutoffs between $1.8$ and $2.5\GeV$, using
several representations for the $\pi\pi$ phase shift and $F_\pi^{V}(s)$ (including estimates of inelastic corrections following~\cite{Schneider:2012ez}) as well as different versions of the conformal polynomial in the $e^+e^-\to3\pi$ fit. Next, the uncertainty in the BL limit is estimated by a fit to singly-virtual data as represented in Fig.~\ref{fig:TFF_spacelike}. Finally, the asymptotic piece~\eqref{asym} is evaluated for $\sm=1.7(3)\GeV^2$, which ensures a smooth matching for $q_1^2=q_2^2=-Q^2$ (note that such a relatively low transition point is indeed expected from light-cone sum rules~\cite{Khodjamirian:1997tk,Agaev:2010aq,Mikhailov:2016klg}). 

\begin{figure}[t]
 \centering
\includegraphics[height=\linewidth,angle=-90,clip]{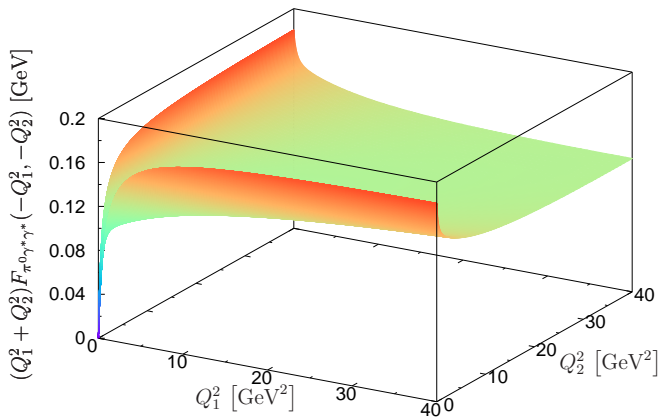}
 \caption{Two-dimensional representation of $(Q_1^2+Q_2^2)F_{\pi^0\gamma^*\gamma^*}(-Q_1^2,-Q_2^2)$, indicating how the respective asymptotic limits are approached.}
 \label{fig:2d}
\end{figure}

In contrast to our framework, analytical approaches to the pion transition form factor in the context of $(g-2)_\mu$, such as vector- and lowest-meson dominance~\cite{Knecht:2001qf,Nyffeler:2016gnb,Husek:2015wta}, rational approximants~\cite{Masjuan:2012wy,Masjuan:2017tvw}, or resonance chiral theory~\cite{Roig:2014uja,Guevara:2018rhj}, are restricted to the space-like region, resulting in an interpolation between the on-shell point $F_{\pi\gamma\gamma}$ and the space-like singly-virtual data. The main advantage of the dispersive approach concerns the fact that all low-energy singularities are reproduced correctly, which enables us to analyze constraints from all low-energy data in a common framework, in particular time-like data from $e^+e^-\to 3\pi$. In this way, the dependence on space-like data for the transition form factor itself is reduced appreciably. 
Indeed, if one were to take the fit to the complete data base at face value, the BL error would shrink to $0.2\times 10^{-11}$ (with a central value increasing to $63.1\times 10^{-11}$), while the generous band chosen in Fig.~\ref{fig:TFF_spacelike} allows us to stay completely agnostic of any of the involved systematics without affecting the overall error estimate too severely.
In addition, within the dispersive approach we can actually predict the doubly-virtual form factor by virtue of its isospin structure~\eqref{Fvs_isospin} and the double-spectral representation~\eqref{low_energy}, thus removing the systematic uncertainty of having to extrapolate singly-virtual data to doubly-virtual kinematics. Finally, the matching scheme that we have developed in this Letter ensures that the asymptotic limits in all directions are correct, see Fig.~\ref{fig:2d}, which is hard to achieve within a resonance model. While our central value lies within the same ballpark as previous calculations, we have therefore, for the first time, provided a robust, comprehensive, and fully data-driven uncertainty estimate.

In general, our error analysis thus shows that, for the pion-pole contribution, the dominant uncertainties are all related to singly-virtual measurements, which points towards
opportunities for future improvements. That is, the normalization error in $F_{\pi\gamma\gamma}$ is likely to shrink by a factor of $2$ at PrimEx-II; the uncertainties in the low-energy region are currently dominated by the systematics of the representation used for the analytic continuation from the time-like $e^+e^-\to 3\pi$ data, which could be reduced by including low-energy space-like data, as forthcoming at BESIII~\cite{Redmer:2018gah}; the uncertainties in the BL limit would be reduced substantially if the Belle data for large virtualities were corroborated~\cite{Kou:2018nap}; and the treatment of the asymptotic region could be improved by comparing to doubly-virtual results from lattice QCD~\cite{Gerardin:2016cqj}. 
Irrespective of such future improvements, our
result~\eqref{result_final} shows that even with currently available
data the pion-pole contribution to HLbL scattering in $(g-2)_\mu$ is
already under very good control, with an uncertainty safely below the
level required for the upcoming Fermilab measurement.

\section*{Acknowledgments}
\begin{acknowledgments}
We thank Johan Bijnens, Antoine G\'erardin, Tobias Isken,
Andreas Nyffeler, Stefan Ropertz, and Peter Stoffer for useful discussions. 
Financial support by
the DFG (CRC 110,
``Symmetries and the Emergence of Structure in QCD''),
the Bonn--Cologne Graduate School of Physics and Astronomy (BCGS),
and the DOE (Grant No.\ DE-FG02-00ER41132)
is gratefully acknowledged.
\end{acknowledgments}

\end{document}